\documentclass[pra,reprint,showpacs,superscriptaddress,nofootinbib]{revtex4-1}

\usepackage{amsbsy}
\usepackage{amsmath}
\usepackage{amsfonts}
\usepackage{amsthm}
\usepackage{color}
\usepackage{mathrsfs}
\usepackage{graphicx}
\usepackage{epstopdf}
\usepackage{dcolumn}
\usepackage{bm}
\usepackage{rotating}
\usepackage{multirow}

 \usepackage{amsmath}    \usepackage[normalem]{ulem}
\usepackage{bm}
\newcommand{\ket}[1]{\vert#1\rangle}

\def\opone{\leavevmode\hbox{\small1\kern-3.8pt\normalsize1}}
 
\usepackage[normalem]{ulem}
\usepackage{bm}

\begin{document}

\title{Performing private database queries in a real-world environment using a quantum protocol}
 
\author{Philip~Chan}
\affiliation{Institute for Quantum Science and Technology, and Department of Electrical \& Computer Engineering, University of Calgary, Canada}
\author{Itzel Lucio-Martinez}
\affiliation{Institute for Quantum Science and Technology, and Department of Physics \& Astronomy, University of Calgary, Canada}
\author{Xiaofan~Mo}
\altaffiliation[Current Address:  ]{Beijing Institute of Aerospace Control Devices, Quantum Engineering Center, China Aerospace Science and Technology Corporation, Beijing 100854}
\thanks{test}
\affiliation{Institute for Quantum Science and Technology, and Department of Physics \& Astronomy, University of Calgary, Canada}
\author{Christoph~Simon}
\affiliation{Institute for Quantum Science and Technology, and Department of Physics \& Astronomy, University of Calgary, Canada}
\author{Wolfgang~Tittel}
\affiliation{Institute for Quantum Science and Technology, and Department of Physics \& Astronomy, University of Calgary, Canada}

\begin{abstract}
In the well-studied cryptographic primitive 1-out-of-$N$ oblivious transfer, a user retrieves a single element from a database of size $N$ without the database learning which element was retrieved.  While it has previously been shown that a secure implementation of 1-out-of-$N$ oblivious transfer is impossible against arbitrarily powerful adversaries, recent research has revealed an interesting class of private query protocols based on quantum mechanics in a cheat sensitive model.  Specifically, a practical protocol does not need to guarantee that database cannot learn what element was retrieved if doing so carries the risk of detection.  The latter is sufficient motivation to keep a database provider honest.  However, none of the previously proposed protocols could cope with noisy channels.  Here we present a fault-tolerant private query protocol, in which the novel error correction procedure is integral to the security of the protocol.  Furthermore, we present a proof-of-concept demonstration of the protocol over a deployed fibre.\end{abstract}

\maketitle

Uncertainty in quantum mechanics can be used to provide security in cryptographic applications, allowing quantum cryptographic protocols to relax the typical assumptions required for security (e.g. an adversary with limited computational power), or even avoid them altogether.  The use of quantum information has proven extremely successful for key distribution,for which quantum key distribution (QKD)\cite{Bennett84,Gisin02,Scarani09} can allow two parties to communicate over a public channel with information theoretic security (i.e. security against an adversary with arbitrarily powerful computational capability, including quantum computers).  The application of quantum information theory to other cryptographic tasks is an interesting topic both because of the insight offered into capabilities of quantum versus classical information coding, and because of the possibility of developing new practical cryptographic protocols with improved security.  Indeed, there are various proposals and experimental demonstrations of quantum cryptographic primitives such as secret sharing\cite{Hillery99,Tittel01}, coin-flipping\cite{Bennett84, Aharonov00, Berlin11}, bit commitment\cite{Ng12, Konig12}, and oblivious transfer (OT)\cite{Giovannetti08, DeMartini09, Jakobi11, Gao11, Konig12}.

When considering cryptographic protocols for deployment, a protocol must ultimately satisfy the following two criteria:
\begin{enumerate}
\item Security:  The protocol must have a rigorous security analysis based on reasonable assumptions about the adversaries.  A strong justification must exist for believing that these assumptions are true.
\item Implementability:  The protocol must be implementable with existing technologies, and must function in the presence of loss and noise (which are inevitable in a realistic implementation).
\end{enumerate}
\noindent However, initially proposed protocols often do not meet both requirements, and in particular often do not consider loss and/or noise in the quantum channel.  Indeed, of the above mentioned protocols, only the bit commitment and OT protocols of ref.~\citenum{Ng12, Konig12} are simultaneously loss- and noise-tolerant, and thus are candidates for real-world implementation.

\begin{table*}[htb]
\centering
\caption{Comparison of the ability of various protocols for private queries to meet the two criteria for deployment (security and implementability).  Note that the cheat sensitive security model may offer the possibility for security with no additional conditions since the impossibility proof\cite{Lo97} may not apply. \label{tab:comparison}}
\begin{tabular}{|c|c|c|c|c|c|}
\hline
~ & ~ & \multicolumn{2}{c|}{Security} & \multicolumn{2}{c|}{Implementability}\\
\cline{3-6}
~ & protocol & \begin{minipage}{0.7in}\center security model\end{minipage} & \begin{minipage}{3in}\center conditions for which security is known to hold\end{minipage} & \begin{minipage}{0.6in}\center loss-tolerant\end{minipage} & \begin{minipage}{0.6in}\center fault-tolerant\end{minipage}\\[1ex]
\hline
\multirow{2}{*}{\begin{minipage}{0.3in}\begin{sideways}\hspace{0.6em}classical\end{sideways} \begin{sideways}information\end{sideways}\end{minipage}} & \begin{minipage}{1in}\center computational\cite{Rabin81}\end{minipage} & standard & \begin{minipage}{2.1in}\vspace{0.3em}adversary has limited classical and quantum computational capability\end{minipage} & N/A & N/A\\[4ex]
& trusted\cite{Naor00, Blundo07} & standard & \begin{minipage}{3in}trusted intermediaries are available\end{minipage} & N/A & N/A\\[4ex]
\hline
\multirow{2}{*}{\begin{minipage}{0.15in}\begin{sideways}quantum information\end{sideways} 
\end{minipage}}
& \begin{minipage}{0.8in}\center noisy-storage\cite{Konig12,Erven13}\end{minipage} & standard & \begin{minipage}{3in}parameters of the adversary's quantum memory (e.g. decoherence as a function of time) are known\end{minipage} & yes & yes\\[1.8ex]
& \begin{minipage}{0.6in}\center GLM\cite{Giovannetti08}\end{minipage} & \begin{minipage}{0.7in}\center cheat sensitive\end{minipage} & \begin{minipage}{3in}no additional conditions\end{minipage} & no & no\\[1.8ex]
& \begin{minipage}{0.7in}\center QKD based\cite{Jakobi11,Gao11} \end{minipage} & \begin{minipage}{0.7in}\center cheat sensitive\end{minipage} & \begin{minipage}{3in}specific attacks discussed in refs.~\citenum{Jakobi11,Gao11}\end{minipage} & yes & no\\[1.8ex]
& \begin{minipage}{0.6in}\center our protocol\end{minipage} & \begin{minipage}{0.7in}\center cheat sensitive\end{minipage} & \begin{minipage}{3in}specific attacks discussed in this work\end{minipage} & yes & yes\\[1.8ex]
\hline
\end{tabular}
\end{table*}

In the case of oblivious transfer, it has been shown that if both parties possess a universal quantum computer it is impossible to simultaneously guarantee that the user, Ursula, can reliably retrieve only a single element while ensuring that the database provider, Dave, has absolutely no knowledge of which element was retrieved\cite{Lo97}.  However this does not mean a practical protocol cannot exist.  First, note that the security criterion allows for reasonable assumptions about the computational capabilities of the dishonest party (e.g. restricting the adversary from having a universal quantum computer).  Indeed, classical OT protocols also rely on one of two assumptions --- that at least some fraction of the intermediaries used to perform the query are trustworthy\cite{Naor00, Blundo07}, or that the adversary has limited classical computational resources\cite{Rabin81}.  In particular, a quantum protocol has been proposed based on the assumption that the adversary has limited noisy quantum storage\cite{Konig12} (which precludes the adversary from possessing a universal quantum computer).  However, new developments (e.g. improvements in computational methods\cite{Kleinjung10,Shor97} or in quantum memory\cite{Lvovsky09, Tittel10, Hammerer10, Simon10, Schindler11, Bussieres13}, respectively) may make these assumptions difficult to justify in the long term.  Second, it may be acceptable in practice to relax security conditions of OT --- that is, one can allow the user to learn more information from the database, and/or the database may be able to gain some information about the query.  Several quantum protocols have been proposed in this vein based on a cheat sensitive model\cite{Giovannetti08, DeMartini09, Jakobi11, Gao11}, in which the database provider is kept honest by the possibility of being caught cheating.  (This type of security can be sufficient if users wish to purchase information privately from a database who spends significant effort gathering and analyzing data, e.g. to make recommendations to investors, as the database must maintain a high quality of service\cite{Jakobi11}.)  In this setting, the protocol need not prevent the database from gaining any information about the user's query, hence protocols may exist in which the assumptions are easier to justify, or in which no assumptions are required at all.  A brief comparison of the properties of the above mentioned protocols for OT and private queries, as well as the protocol we present in this work is given in Table~\ref{tab:comparison}, and we review these protocols in further detail in the Supplementary Information.

In this work, we propose a private query protocol based on the protocols of ref.~\citenum{Jakobi11,Gao11}, retaining the advantages of those works while addressing the remaining obstacle to meeting the implementability criterion.  This is accomplished using a novel error correction algorithm, in which the algorithm and its associated parameters are tailored to provide the desired level of security in the private query protocol.  Furthermore, we note that the novel error correction procedure used to provide fault-tolerance also provides additional opportunities for Ursula to verify Dave's honesty, thus enhancing the cheat sensitive property of the protocol.  Hence, we show that error correction is not simply necessary to meet the implementability criterion, but is integral to the security criterion as well.

\section{Results\label{sec:results}}

As in ref.~\citenum{Jakobi11, Gao11}, we implement a cheat sensitive private query protocol based on the SARG04 Quantum Key Distribution (QKD) protocol\cite{Scarani04}.  The functionality of the protocol can be described as implementing probabilistic $n$-out-of-$N$ OT --- that is, Ursula will, on average, learn the value of $\bar{n}$ bits (where $\bar{n}$ is small) of the database with high confidence (for brevity, we often simply describe such bits as being known to Ursula).  She will also have probabilistic knowledge of other bits of the database (i.e. she can guess their value with better than 50\% probability).  In this scheme, a private query on an $N$-bit database is made possible using an $N$-bit oblivious key (for simplicity, we consider each element of the database to be a single bit) generated by the quantum protocol, in which the goal is to ensure that Ursula knows, on average, $\bar{n}$ bits of the oblivious key, whose positions are unknown to Dave.  In the following sections, we give a description of the protocol for generating an oblivious key and using it to perform private queries, give an overview of the error correction procedure, and then conclude with a brief discussion on security.

\subsection{Description of the protocol}

A detailed description of the honest protocol for performing a private query is as follows (see Figure~\ref{fig:protocol} for a graphical representation of the protocol):

\begin{figure*}[bhtp]
\centerline{\includegraphics[width=40pc]{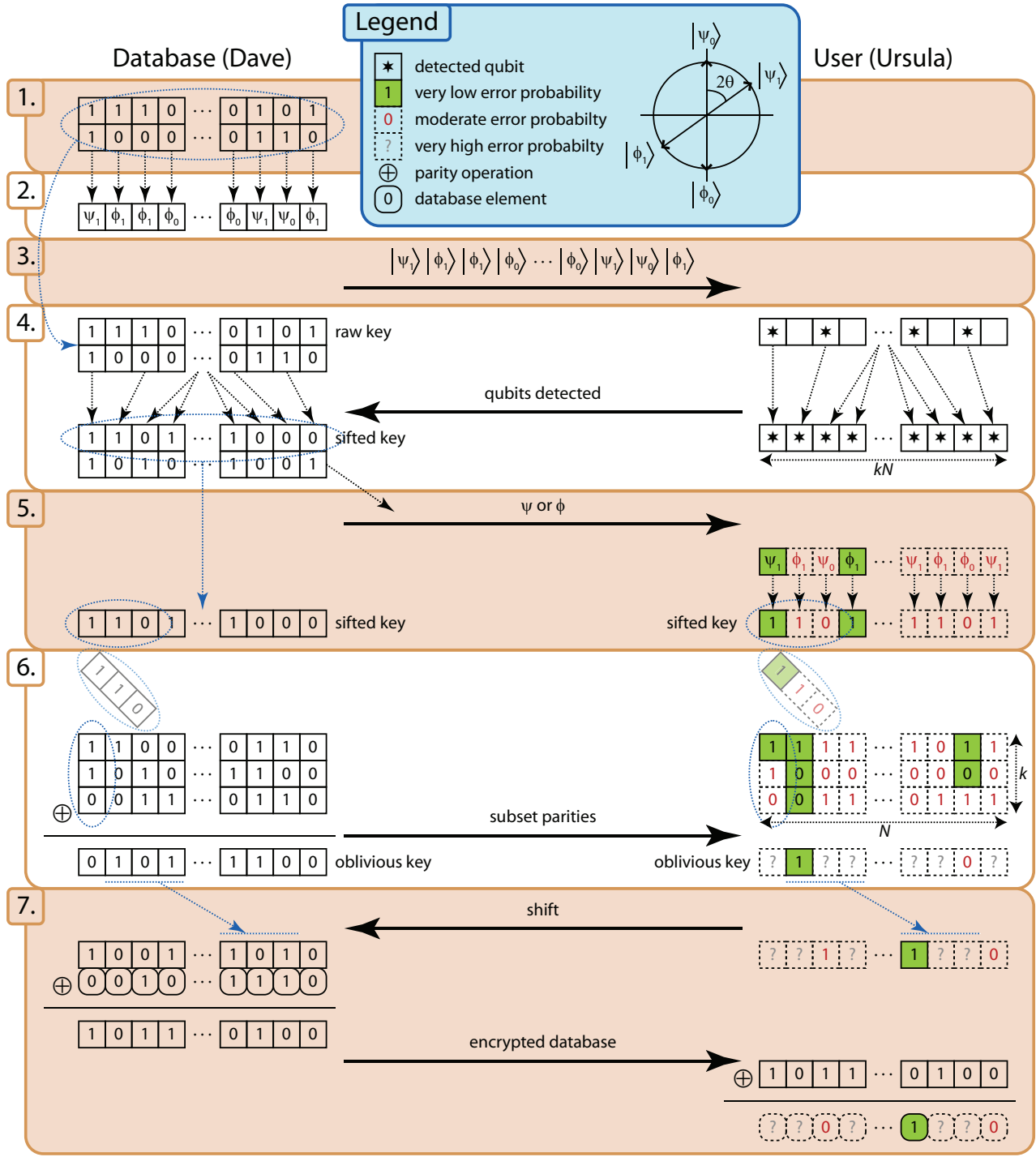}}
\caption{Graphical representation of the private query protocol.  The steps indicated on the left margin correspond to the steps described in the text.\label{fig:protocol}}
\end{figure*}

\begin{enumerate}
\item Dave generates two long strings of classical bits uniformly at random, and records their values.  Each string should be $\approx\frac{kN}{t}$ bits in length, where $k$ is a parameter determined by the previously agreed-upon error correction procedure (to be discussed later), $N$ is the length of the database, and $t$ is the transmission of the link between Ursula and Dave.
\item Dave uses each pair of classical bits generated above to choose a quantum state from a set of four previously agreed upon non-orthogonal states (shown in Figure~\ref{fig:protocol}), and prepares qubits accordingly.  A random bit from the first string determines whether the state is prepared in the $0$-basis (spanned by $\ket{\psi_0}$ and $\ket{\phi_0}$) or the $1$-basis (spanned by $\ket{\psi_1}$ and $\ket{\phi_1}$), and the corresponding random bit in the second string determines whether the $\psi$ or $\phi$ state in each basis is chosen.  The first random string forms Dave's raw key, for which the bit values correspond to the bases in which he prepared the qubits.
\item Dave sends the qubits encoded into single photons to Ursula using a possibly lossy and noisy quantum channel.
\item Ursula makes projection measurements using either the $0$- or $1$-basis, chosen uniformly at random, and records the measurement bases and the results.  Ursula publicly announces the cases in which she detected a photon, and Ursula and Dave both discard all the events in which Ursula failed to detect the photon. The protocol proceeds to the next step once Ursula has succeeded in detecting $kN$ photons.  Dave keeps the corresponding $kN$ bits from his raw key to form his sifted key.
\item Dave publicly announces his second string of random bits (used to select whether he encoded the qubits into a $\psi$ or $\phi$ state), which, combined with knowledge from Ursula's measurements (and, for the moment, assuming a noiseless channel), allows her to conclusively identify whether the state was encoded in the $0$- or $1$-basis with probability $p_\mathrm{c} = \frac{\sin^2(\theta)}{2}$.  Note that when Ursula's measurements yielded inconclusive results, which occurs with probability $p_\mathrm{i} = 1 - p_\mathrm{c}$, she gains probabilistic information about the basis.  This information can be quantified by the probability that she incorrectly identifies the basis, $e_\mathrm{i} = \frac{\cos^2(\theta)}{1 + \cos^2(\theta)}$.  A noisy channel will affect the probabilities $p_\mathrm{c}$, $p_\mathrm{i}$, and $e_\mathrm{i}$, as well as result in a non-zero error rate for conclusive measurements, denoted $e_\mathrm{c}$.  Like Dave, Ursula associates classical bit values to the quantum states based on the basis, and forms her sifted key using the most likely values of the bits given her measurement results.
\item Dave divides his sifted key into $N$ $k$-bit blocks, and computes each bit of his oblivious key as the parity of the $k$ bits in each block (the parity is 0 if an even number of the $k$ bits is 1, and 1 otherwise).  He publicly announces which bits form each block.  In addition, according to a previously agreed upon error-correcting code, he also sends the parities of several subsets of the $k$ bits to Ursula.  Using this information, along with her sifted key and knowledge of whether the measurements were conclusive or inconclusive, Ursula computes the most likely value of each oblivious key bit, as well as the probability that this value is incorrect, denoted $e_\mathrm{k}$.  The error-correcting code is selected such that Ursula will only have a high confidence (or low $e_\mathrm{k}$) in $\bar{n}$ bits on average, where $\bar{n}$ is typically a few bits.  If Ursula does not learn any bits of the protocol (due to its probabilistic nature), the protocol must be restarted.
\item Ursula selects a shift value that aligns one of the bits she knows in the oblivious key to the bit in the database that she wants to know.  She communicates this shift value classically to Dave, who applies the shift to his oblivious key, and then uses it to encrypt the database using the one-time-pad\cite{Vernam26}.  He then sends the encrypted database to Ursula, who can only decrypt the bits for which she knows the corresponding oblivious key bit.  If Ursula knows multiple bits of the oblivious key she will learn multiple bits of the database.  However, the shift only allows her to select the location of a single bit of the database, with the remaining learned bits distributed randomly.
\end{enumerate}

\subsection{Error-correcting codes for private queries}
Let us now examine step 6 of the protocol in more detail.  Our error correction procedure (see Supplementary Information for a full description) is inspired by syndrome decoding of error-correcting codes such as Hamming codes\cite{MacKay03}, which can operate on a few bits at a time.  However, it is important to note that in the context of private queries error correction is integral to determining how much information Ursula learns about the oblivious key, creating unique requirements that made it necessary to investigate and design novel error-correcting codes and error correction procedures.  In particular, the goal when designing an error-correcting code for private queries is not to simply maximize the probability of successful decoding as it is in standard communications applications.  Rather, a specific success probability is desired in order to ensure that Ursula only learns a few bits of the oblivious key.  Furthermore, to prevent Ursula from learning a large amount of probabilistic information about the remaining bits of the key, it is desirable to keep $e_\mathrm{k}$ as high as possible in those cases in which decoding does not succeed.

In addition there are two main technical differences between error correction in private queries and in communications.  First, note that in order to recover the value of the oblivious key bit, Ursula need only determine the parity of the $k$-bits, and not the individual values of the $k$ bits as would typically be the case for error correction.  Hence, the error correction procedure seeks the most likely parity of the $k$-bit block, and successful decoding does not depend on having a high probability of identifying the correct values of the $k$-bit block as long as it is possible to identify whether an even or odd number of errors occurred.  Second, the input bits can be divided into those with low error rate (conclusive measurements), and those with very high error rate (inconclusive measurements).  We note that it is the interaction of this latter property with the short block lengths used ($k \le 10$) that allows uncertainty to be maintained after error correction, thereby limiting the amount of information that Ursula learns about the database.

The error-correcting codes used in this work are tailored based on the experimental parameters (i.e. conclusive and inconclusive probabilities, $p_\mathrm{c}$ and $p_\mathrm{i}$ and the associated error rates $e_\mathrm{c}$ and $e_\mathrm{i}$) in order to achieve the goals discussed above.  In order to quickly evaluate error-correcting codes, we define two thresholds, $t_\mathrm{U}$ and $t_\mathrm{D}$.  When $e_\mathrm{k} \le t_\mathrm{U}$, Ursula considers the oblivious key bit to be known.  When $e_\mathrm{k} \le t_\mathrm{D}$, Dave considers Ursula to have significant partial information about that bit.  These thresholds should be selected based on the requirements of the application.  In this work, we use $t_\mathrm{U} = 10^{-3}$ and $t_\mathrm{D} = \frac{1}{3}$.  In order to reduce the probability of error in Ursula's oblivious key bit below her threshold (i.e. $e_\mathrm{k} \le t_\mathrm{U}$), the error correction process must sufficiently reduce $e_\mathrm{k}$ when her quantum measurements succeeded in obtaining a large amount of information about the $k$ bits (i.e. when most or all measurements were conclusive).  However, the error correction will also reduce $e_\mathrm{k}$ if several measurements were inconclusive.  Hence, the error rate for inconclusive measurements, $e_\mathrm{i}$, is of particular importance to the fraction of bits for which $e_\mathrm{k} \le t_\mathrm{D}$.  With this in mind, a smaller angle between states (characterized by $\theta$ as shown in Figure~\ref{fig:protocol}) has, in addition to those benefits noted in ref.~\citenum{Gao11} (i.e. reduced quantum communication, improved database security, and better control over the number of bits Ursula learns), the benefit of reducing the partial information from inconclusive measurements.  However, there is a trade-off between these benefits and the fact that the error rate for conclusive measurements is also increased due to a reduced signal-to-noise ratio, making it more difficult to achieve $e_\mathrm{k} \le t_\mathrm{U}$.  A detailed description of the selection of our error-correcting codes is given in the Supplementary Information.

\subsection{Security of the protocol}

Let us now discuss how the steps in the above protocol contribute to security, beginning with a discussion of user privacy.  User privacy is protected by the cheat sensitive property of the protocol, which allows a dishonest database to be detected.  This property stems from step 4 of the protocol as Ursula randomly selects between two possible (non-commuting) measurements and does not announce which measurement she performed.  Her security thus stems from the complementarity principle as her interpretation of her measurement results is dependent on her choice of measurement basis, with the protocol designed such that the classical bit value she assigns to each result is perfectly correlated with her basis choice (see step 5 and the Supplementary Information for more details).  In the case that Dave is honest (and for the moment, assuming a noiseless system), Ursula's classical bit values for conclusive measurements will also be perfectly correlated with the classical bit values Dave used to select which quantum states he encodes.  If Dave is dishonest, and supposing he can send a state such that Ursula's measurement is conclusive regardless of which measurement basis she chooses (a realistic attack is analyzed in the Supplementary Information), Ursula's interpretation of her measurements remain unchanged, hence her classical bit values are still perfectly correlated to her choice of basis.  Since this choice is never revealed to Dave, he does not know which bit value she obtains.  This leads to the cheat sensitivity in the protocol, as the dishonest database may be detected during error correction (since he sends parity values uncorrelated with Ursula's classical bit values), or after completion of the protocol since he may send incorrect query results.  Furthermore, note that the error correction procedure in step 6 only involves one-way communication from Dave to Ursula, hence Dave gains no information regarding the results of the error correction procedure.

On the other hand, Ursula's limited knowledge about the oblivious key stems from the superposition principle in quantum mechanics.  Specifically, note that in step 2 Dave prepares qubits in non-orthogonal states, and that Ursula can thus not deterministically distinguish between these states.  As such, Ursula's measurements only give her limited information, even after Dave reveals some information about which state he sent in step 5.  Furthermore, note that Ursula must declare which bits were lost during transmission (or detection) in step 4, prior to receiving classical information indicating whether a $\psi$ or $\phi$ state was sent.  This makes the protocol loss-tolerant while ensuring that Ursula cannot choose which bits to keep based on whether her measurements were conclusive or inconclusive, even if she uses a heralded quantum memory to delay her measurements until after step 5.  Note that in step 6, Ursula does have the ability to restart the protocol if the results are unfavorable as Dave cannot verify whether she indeed learned no bits of the oblivious key.  However, choosing an error-correcting code such that $\bar{n}$ is a few bits ensures that the probability for Ursula to not know any bits is very low, and allows Dave to abort the protocol after a small number of declared failures by Ursula (preventing her from repeatedly declaring failure until she obtains a very favorable result).

Furthermore, a dishonest user may gain an advantage by deviating from the honest protocol.  It has been shown that Ursula could perform an unambiguous state discrimination (USD) measurement\cite{Herzog05, Raynal06} in order to slightly improve her probability of conclusive measurements, which allows her to learn a few additional bits of the oblivious key\cite{Jakobi11}.  However, this comes at the expense of gaining no information about the bit value (i.e. $e_\mathrm{i}=0.5$) when the USD measurement gives inconclusive results.  While this probabilistic information was not previously considered useful\cite{Jakobi11, Gao11}, it is an important input to the error correction process.  Thus, the effectiveness of this attack is reduced in the presence of error correction, and our analysis in the Supplementary Information shows that in some cases performing a USD measurement actually reduces the number of bits of the oblivious key that Ursula learns as compared to the honest measurements.  Note that only individual USD measurements have been considered, and coherent attacks (e.g. an optimized USD measurement on the $k$ qubits that form each oblivious key bit) remain an interesting open question.

We also note that Ursula and Dave are adversarial in nature in the protocol, and thus may not cooperate when estimating the error rate in order to select an appropriate error-correcting code.  An error-correcting code that is not well suited to the actual error rate in the system will either result in Ursula learning too few or too many bits of the oblivious key, but does not impact user security.  Hence the database does not have any motivation to falsify the error rate, but the user would like the database to think the error rate is larger than it is in reality, leading to the selection of an error-correcting code that gives her more information.  In our analysis (detailed in the Supplementary Information), we find that Dave can ensure that he has a reasonable level of security by determining the error rate of devices under his control (potentially by intentionally introducing noise) and selecting an error-correcting code accordingly.  In addition, even if Ursula's devices introduce some additional error that Dave does not account for in his security analysis, the protocol is still successful for her.

\subsection{Experimental and simulated performance of our protocol}

\begin{figure}[thbp]
\centerline{\includegraphics[width=\columnwidth]{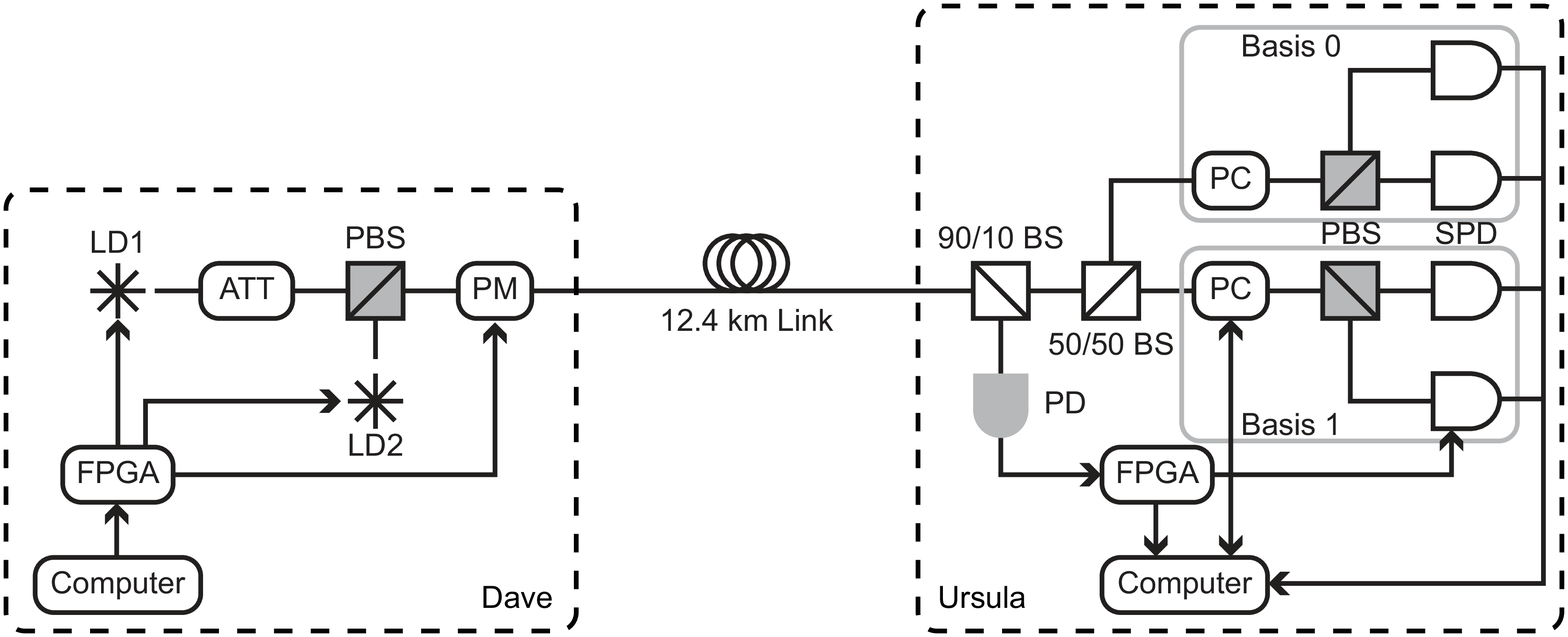}}
\caption{Diagram of the experimental setup.  The database (Dave) uses a computer and field-programmable gate-array (FPGA) to control the generation of polarization qubits via an attenuated laser diode (LD1 and ATT) and polarization modular (PM).  Quantum frames\cite{Lucio09} (sequences of strong light for timing and stabilization) are generated by a second laser diode (LD2) and merged using a polarizing beam-splitter (PBS).  Light is transmitted from Dave to Ursula through a 12.4~km dark fiber link with 4.5~dB loss between SAIT Polytechnic and the University of Calgary.  Ursula splits off 10\% of the incoming light (90/10 BS) to a photodiode (PD) used to detect the quantum frames.  The 50/50 BS is used to passively select a random measurement basis.  The apparatus for each basis consists of a polarization controller (PC), a PBS, and two single photon detectors (SPD) to make the projection measurement.  Upon detecting a quantum frame, Ursula's FPGA triggers the SPDs and initiates data collection by the computer, or polarization compensation, as appropriate.\label{fig:setup}}
\end{figure}

\begin{table}[htb]
\centering
\caption{Parameters for the private query protocol as measured in our experiment with standard detectors, and simulated for low-noise detectors.  The value of $\theta$ (including standard deviation) is measured using classical light.  For the probabilities of conclusive measurements, $p_\mathrm{c}$, and error rates for conclusive and inconclusive measurements, $e_\mathrm{c}$ and $e_\mathrm{i}$, the standard error expected based on Poissonian counting statistics for the $10^7$ bits in each query is negligible compared to the observed variations across the queries performed.  The observed standard deviations are attributed to time-varying error in the alignment of the measurement bases at the receiver as a result of channel instability.  Note that the measurement results for the $\mu=9.5 \pm 0.47$ case show more variation in the parameters than for the $\mu=0.95 \pm 0.047$ case due to short-term fluctuations that are averaged out by the long data collection time needed to acquire the $10^7$ bits per query in the $\mu=0.95 \pm 0.47$ case.\label{tab:params}}
\begin{tabular}{|c|c|c|c|}
\hline
~ & \multicolumn{2}{c|}{standard detectors} & low-noise detectors \\
\hline
$\mu$ (photons) & $0.95 \pm 0.047$ & $9.5 \pm 0.47$ & 1 \\
$\theta$ ($^\circ$) & $35.6 \pm 0.49$ & $35.6 \pm 0.49$ & 25 \\
$p_\mathrm{c}$ (\%) & $16.1 \pm 0.29$ & $16.1 \pm 0.93$ & 9.22 \\
$e_\mathrm{c}$ (\%) & $4.4 \pm 0.59$ & $4.6 \pm 0.38$ & 1.91 \\
$e_\mathrm{i}$ (\%) & $41.24 \pm 0.08$ & $41.3 \pm 0.64$ & 45.12 \\
$k$ (bits) & 10 & 10 & 9 \\
\hline
\end{tabular}
\end{table}

\begin{table*}[bht]
\centering
\caption{Experimental and simulated results for the quantum private queries.  The following figures of merit are used:  the average number of bits learned by the user per query, $\bar{n}$, the average proportion of the database where the user has significant partial information (i.e. $e_\mathrm{k}\le t_D$), $\bar{m}$, and the failure probability (i.e. that the user learns zero bits), $P_0$.\label{tab:results}}
\begin{tabular}{|c|c|c|c|c|c|}
\hline
~ & \multicolumn{2}{c|}{$\mu=0.95 \pm 0.047$} & \multicolumn{2}{c|}{$\mu=9.5 \pm 0.47$} & low-noise \\
\cline{2-6}
~ & experimental & simulated & experimental & simulated & simulated \\
\hline
$\bar{n}$ (bits) & $4.1 \pm 2.4$ & $3.2 \pm 1.1$ & $3.9 \pm 3.1$ & $3.5 \pm 1.9$ & 4.35 \\
$\bar{m}$ (\%) & $6.1 \pm 0.25$ & $6.1 \pm 0.25$ & $6.3 \pm 1.4$ & $6.3 \pm 1.3$  & 0.96 \\
$P_0$ (\%) & $9.1 \pm 9.1$ & 8.8 & $8.7 \pm 2.9$ & 9.4 & 1.29 \\
\hline
\end{tabular}
\end{table*}

\begin{figure*}[bth]
\centerline{\includegraphics[width=36pc]{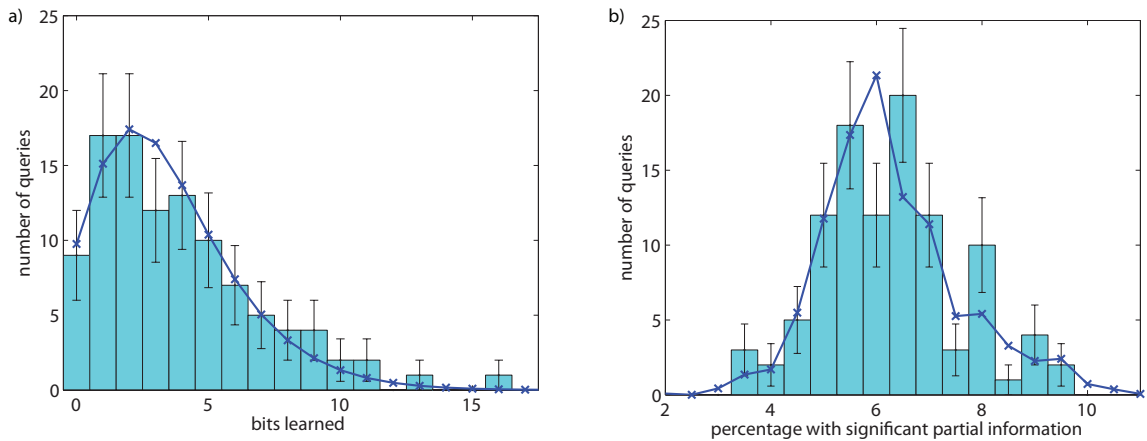}}
\caption{Histograms for the information gained by the user in the 104 queries performed in the $\mu=9.5 \pm 0.47$ case.  a) The number of bits learned by the user.  b)  The percentage of the database of which the user learns significant partial information.  In both figures error bars for the experimental results represent one standard deviation assuming Poissonian counting statistics, and the blue crosses show the expected distribution obtained from Monte Carlo simulations.\label{fig:results}}
\end{figure*}

The experimental and simulated results for these codes are shown in Table~\ref{tab:results}.  The simulated results corresponding to our experiment are derived from Monte Carlo simulations taking into account the variation in the parameters shown in Table~\ref{tab:params}.  Figure~\ref{fig:results} compares the distribution of the results over the 104 queries performed in the $\mu=9.5 \pm 0.47$ case with the simulation results, showing good agreement between the two.  Note that in both experimental cases, no errors were observed in the bits learned by Ursula (i.e. for which $e_\mathrm{k} \le 10^{-3}$), with a total of 45 bits learned in 11 queries when $\mu=0.95 \pm 0.047$ and 405 bits learned in 104 queries when $\mu=9.5 \pm 0.47$.

We performed an experimental demonstration of private queries over a 12.4 km fiber link between the University of Calgary and SAIT Polytechnic, using our BB84\cite{Bennett84} QKD system\cite{Lucio09} (with a small modification to the hardware to set $\theta=35.6^\circ\pm 0.49^\circ$ --- all other differences between our protocol and BB84 QKD are in the classical post-processing).  Our experimental setup is shown in Figure~\ref{fig:setup} (see ref.~\citenum{Lucio09} for a detailed description).  Note that our demonstration uses weak coherent pulses rather than single photons, and hence database privacy requires the assumption that Ursula is not able to exploit pulses containing multiple photons (adapting the protocol for weak coherent pulses, e.g. using decoy states as in QKD\cite{Hwang03,Wang05,Lo05,Wehner10}, remains an open question, and we discuss this possibility further in the Supplementary Information).  We consider a database size of $N=10^6$ and, based on measured error rates for our system, an error-correcting code with $k=10$ was selected, thus requiring $10^7$ measured qubits per query.  Note that we did not consider $k>10$ due to computational constraints when searching for the best possible construction of the error-correcting code.  A total of 11 queries was performed using a mean number of photons per pulse of $\mu=0.95 \pm 0.047$ to show that the protocol can function at the single photon level.  In this setting, our system took approximately 4.5 hours to accumulate the $10^7$ bits of data needed for one private query.  In order to quickly collect statistics, we repeated the experiment with mean number of photons per pulse increased to $\mu=9.5 \pm 0.47$, performing 104 queries.  While the multi-photon emissions at this $\mu$ are likely to compromise the security of the protocol if Ursula monitors the pulses outside Dave's laboratory, this value corresponds to $\sim0.95$ photons per pulse at the detectors, ensuring that multi-photon detection events do not skew the detection statistics.  The measured parameters that determine the performance of the protocol are shown in Table~\ref{tab:params} (note that the experimentally measured parameters at both mean photon numbers are the same to within one standard deviation), along with parameters for a theoretical simulation of what could be achieved using state-of-the-art detectors\cite{Marsili13, Yan12}.  These detectors allow for significantly reduced noise (they feature dark count rates $\approx 100$~Hz), and, in the case of ref.~\citenum{Marsili13}, detection efficiencies up to 93\%.  With the improved signal-to-noise ratio, we select the parameters of the protocol to be $\theta=25^\circ$ and $k=9$.

In addition, our simulation results show that the primary obstacle to improving database security in the protocol is noise in the system, which can be greatly reduced by state-of-the-art single photon detectors.  These detectors can also improve the rate at which queries can be performed by almost an order of magnitude because of their higher detection efficiencies.  Further improvement of this rate is straightforward, as QKD systems can easily be adapted to perform this protocol.  A state-of-the-art BB84 QKD system has shown that data can be accumulated at a rate of $10^6$ to $10^7$ bits per second, depending on the distance between Ursula and Dave\cite{Dixon08}.  For the parameters in our experimental demonstration, this would allow one private query to be performed every few seconds.  The amount of data required can also be reduced by repeating a short oblivious key over a longer database and then applying a shift as before to allow Ursula to select the desired bit.  This would allow queries to be performed more often, or equivalently, allow queries to be performed on a larger database in the same amount of time.  However, this comes at the expense of database security, as the user is able to learn additional bits for each repetition of the key (though not in locations of her choice, as only a single shift value is communicated).  We also note that a modification to the protocol of ref.~\citenum{Jakobi11} has recently been proposed that reduces the amount of quantum communication required\cite{PandurangaRao13}, however applying this modification to our protocol is not straightforward.

\section{Discussion}

We have proposed and demonstrated, over deployed optical fibres, a quantum protocol for private queries using the cheat sensitive model.  This first demonstration of private queries in a real-world setting was made possible by the development of a protocol which integrates a novel error correction procedure.  Our analysis of this protocol has shown that error correction plays a pivotal role in the security, both in terms of controlling how much information the user learns, and in providing the ability for Ursula to detect a dishonest database.  While our security analysis is currently limited to several specific attacks, it is important to note that the error correction should be viewed as an important tool for tailoring the amount of information learned by the user, and hence may be adaptable to a more general scenario where Ursula makes more powerful measurements.  In this general view, database security stems from the fact that quantum mechanics allows a protocol to be designed where the user cannot extract full information about the quantum states sent, and error correction allows the extracted information to be processed into an oblivious key with the desired distribution of information for private queries.  Furthermore, quantum mechanics allows such a private query protocol to be set up such that the correlation between Ursula and Dave's classical raw key bits is destroyed if Dave can control which bits of the oblivious key Ursula learns.  Hence, the methods presented in this work should provide a strong basis for the further development of cheat sensitive quantum protocols.

\section{Acknowledgments}The authors thank M. Jakobi, M.V. Panduranga Rao and C. Erven for useful discussions, V. Kiselyov for technical support, SAIT Polytechnic for providing laboratory space, and acknowledge funding by NSERC, QuantumWorks, General Dynamics Canada, iCORE (now part of AITF), AITF, CFI, and AAET.

\onecolumngrid

\section{Supplementary Information}

\section{Review of Oblivious Transfer and Private Queries}

An ideal 1-out-of-$N$ oblivious transfer protocol simultaneously guarantees that (a) that the user, Ursula, is able to retrieve a single element from the $N$-bit database, and (b) that the database provider, Dave, cannot gain any information about which element was retrieved.  However, it has been shown that, assuming a universal quantum computer, if a protocol meets condition (b) then condition (a) implies that Ursula can access every element of the database\cite{Lo97}.  As such, it is impossible for a protocol to implement ideal oblivious transfer without making assumptions in the security model.  Alternatively, the class of protocols we refer to as private queries avoids the impossibility proof by implementing functionality similar to 1-out-of-$N$ OT.  Such protocols offer a reduced level of privacy up front, but this reduction in privacy may allow secure protocols using assumptions that are easier to justify, or in which no assumptions are required at all.  In this section, we briefly review protocols for oblivious transfer and private queries.\cite{Lo97}

In classical information theory, protocols for OT rely on one of two assumptions --- that at least some fraction of the intermediaries used to perform the query are trustworthy\cite{Naor00, Blundo07}, or that the adversary has limited classical computational resources\cite{Rabin81}.  The former assumption can be difficult to assess, as one must both believe that the intermediaries will not collude with each other, and that their infrastructure is secured against attacks.  The latter assumption is shared with today's public key cryptography infrastructure, and is hence well justified in the short term.  However, in the long term, the security of such systems can be compromised by advances in algorithms (e.g. ref.~\citenum{Kleinjung10}) or hardware such as a quantum computer\cite{Shor97}.

A quantum 1-out-of-2 OT protocol has also recently been proposed using the noisy-storage model\cite{Konig12}, where it is assumed that the dishonest party has a limited ability to store quantum information, and that the amount of information that can be faithfully stored decreases over time due to noise in the quantum memories (note that this protocol is loss- and fault-tolerant, as quantum memories are not required by the honest protocol).  Since quantum memories are a basic component in a universal quantum computer, this assumption means that the proof that ideal OT is not possible\cite{Lo97} does not apply.  Thus, perfect privacy is possible under this model, and this has indeed been shown in the protocol of ref.~\citenum{Konig12}.  An experimental demonstration of the protocol has also recently been performed\cite{Erven13}, showing that it meets the implementability criterion.  As with the classical OT protocols relying on assumptions about the adversaries computational capabilities, this assumption is well justified in the short term given current quantum memories.  However, there is no fundamental principle limiting the adversaries ability to store quantum information, and recent advances in quantum memories\cite{Lvovsky09, Tittel10, Hammerer10, Simon10, Schindler11, Bussieres13} threaten the validity of this assumption in the long term.

The private queries approach to OT using cheat sensitivity was first proposed in ref.~\citenum{Giovannetti08}.  This protocol does not satisfy condition (b) above, since a dishonest database could gain complete information about which element Ursula retrieved.  However, the protocol still offers security for Ursula as she has, in principle, the potential to detect Dave's attempt to gain information about her query, thus discouraging Dave from cheating.  Note that condition (a) was also not satisfied, as a dishonest user could sacrifice her ability to verify Dave's honesty in order to obtain a small number of additional elements (although, this is not a significant loss of privacy for the database if $N$ is large).  An experimental proof-of-principle demonstration of this protocol was subsequently performed\cite{DeMartini09}, however, as Dave could hide his attempts to cheat if there was significant transmission loss and/or errors in the quantum channel, the protocol is not practical under realistic conditions.  Ref.~\citenum{Jakobi11} then proposed a probabilistic $n$-out-of-$N$ OT protocol based on the SARG04 Quantum Key Distribution (QKD) protocol\cite{Scarani04}, which was then generalized\cite{Gao11}.  This protocol allows Dave to gain information about Ursula's query, but only at the risk of introducing errors into the element Ursula retrieved, thereby allowing a dishonest database to be detected.  The protocol also did not satisfy condition (a) above as Ursula gains probabilistic information about elements of the database she does not request.  Interesting features of this protocol are the ability to tolerate loss in the channel, as well as the fact that it is simple to implement using existing QKD technology.  However, noisy channels were left as an open question, preventing implementation of the protocol in realistic scenarios.  Finally, our protocol proposed in this work represents the first cheat sensitive protocol to be both loss- and fault-tolerant, making it suitable for implementation in a realistic environment.

\section{Quantum State Identification}

\begin{figure}[thbp]
\centerline{\includegraphics[width=9pc]{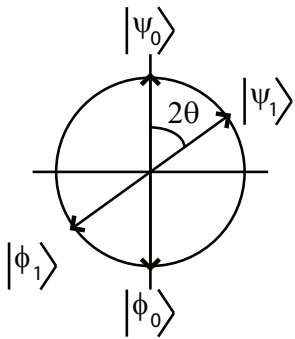}}
\caption{Quantum states used in the private query protocol shown on a plane of the Bloch sphere.\label{fig:SI_states}}
\end{figure}

In our protocol, the database provider, Dave, encodes each qubit into one of four randomly chosen quantum states, $\left\lvert\psi_0\right\rangle$, $\left\lvert\psi_1\right\rangle$, $\left\lvert\phi_0\right\rangle$ or $\left\lvert\phi_1\right\rangle$, as shown in Figure~\ref{fig:SI_states}.  The user, Ursula, measures each qubit in either the 0-basis, spanned by $\left\lvert\psi_0\right\rangle$ and $\left\lvert\phi_0\right\rangle$, or the 1-basis, spanned by $\left\lvert\psi_1\right\rangle$ and $\left\lvert\phi_1\right\rangle$.  After these measurements, Dave tells Ursula whether each qubit was encoded into one of the $\psi$ states or one of the $\phi$ states.  In order to demonstrate the state identification process, suppose Ursula measured in the 0-basis, and Dave declares that he sent one of the $\psi$ states.  If Ursula's measurement result was $\left\lvert\phi_0\right\rangle$, she knows Dave could not have sent $\left\lvert\psi_0\right\rangle$ as these two states are orthogonal.  Hence Dave must have sent $\left\lvert\psi_1\right\rangle$.  This is a conclusive result, and occurs with probability $p_\mathrm{c}=\frac{\sin^2(\theta)}{2}$.  Alternatively, if Ursula's measurement result was $\left\lvert\psi_0\right\rangle$, she only knows that the state was more likely to have been $\left\lvert\psi_0\right\rangle$ than $\left\lvert\psi_1\right\rangle$.  This is an inconclusive result, occurring with probability $p_\mathrm{i} = 1-p_\mathrm{c}$.  As the two potential states are associated with different classical bit values (as indicated by the subscripts), Ursula only gains probabilistic knowledge from this measurement result.  This corresponds to an error rate of $e_\mathrm{i} = \frac{\cos^2(\theta)}{1+\cos^2(\theta)}$ in the ideal case (i.e. when no other sources of error are present).

Let us now examine how this state identification process leads to user privacy, considering first the honest protocol.  In the above example where Dave sent one of the $\psi$ states and Ursula measures in the 0-basis, note that Ursula can only get a conclusive measurement result if Dave sent the $\ket{\psi_1}$ state.  If Ursula instead measures in the 1-basis, she can only get a conclusive measurement if Dave sent the $\ket{\psi_0}$ state.  Hence, for any given qubit that Dave sends, Ursula's choice of measurement determines whether a conclusive result is possible --- she never gets a conclusive result if she measures in the same basis in which Dave encoded the qubit.  Since she never reveals her choice of measurement basis to Dave, he cannot know which of her measurements gave conclusive results.

Now, let us consider the case in which Dave is dishonest.  In this case, Dave wishes to break the correlation between Ursula's choice of measurement basis and the conclusiveness of her measurement results.  Ideally, he would like to choose whether Ursula will get a conclusive or inconclusive measurement result, regardless of which measurement she makes.  For ease of discussion, we assume here that Dave can send a quantum state that accomplishes this goal (we discuss a more realistic attack in Section~\ref{sec:cheating}).  Since Ursula is honest, she makes the same measurements as before, and interprets them assuming Dave is honest.  In the above example, in which Dave declares he sent one of the $\psi$ states, if Ursula measures in the 0-basis, she will either conclusively identify that Dave sent the $\ket{\psi_1}$ state, or inconclusively identify that Dave likely sent the $\ket{\psi_0}$ state.  If she instead measured in the 1-basis, she will either conclusively identify that Dave sent the $\ket{\psi_0}$ state, or inconclusively identify that Dave likely sent the $\ket{\psi_1}$ state.  Recall that the classical bit values that form the raw keys in the protocol are given by the basis of the state that Ursula believes Dave sent (and correspond to the subscripts in the ket notation).  Thus, Ursula's raw key bits are anti-correlated with her choice of measurement basis for conclusive results, and correlated for inconclusive results.  Hence, if Ursula's choice of measurement basis does not determine whether a measurement is conclusive, it instead determines her raw key bits.  In this case, since she never reveals her choice of measurement basis, Dave cannot know her raw key bits.  This leads to the cheat sensitivity in the protocol as the fact that Dave has no knowledge of Ursula's raw key bits may be detected during error correction, and if not detected, results in incorrect query responses.  A more detailed analysis of the cheat sensitivity is given in Section~\ref{sec:cheating}.

\section{Error Correction}

We use a parity-based forward error-correcting code operating on $k$-bit blocks (corresponding to the $k$ bits used to compute one oblivious key bit), where Dave sends the parity of several subsets of the $k$ bits to Ursula.  The construction of the code is normally described as a parity check matrix, denoted $\mathbf{H}$, and is known to both Ursula and Dave.  The parity computation for the $j^\mathrm{th}$ oblivious key bit is then given by:

\begin{equation}
\vec{p}_j = \mathbf{H}\vec{d}_j\pmod 2\label{eqn:parity}
\end{equation}

\noindent where $\vec{p}_j$ is a vector of computed parity bits (which Dave sends to Ursula) and $\vec{d}_j$ is a vector containing the $k$ bits that Dave uses to compute a single oblivious key bit.  For each oblivious key bit, Ursula has a corresponding $k$-bit vector, $\vec{u}_j$, in which each bit stems from a conclusive or an inconclusive measurement that have, respectively, error rates of $e_\mathrm{c}$ and $e_\mathrm{i}$.  Ursula can estimate these error rates over the entire protocol by comparing the parities, $\vec{p}_j$, she receives from Dave and the parities she computes locally using $\vec{u}_j$.  Using these error rates, Ursula's error correction procedure for each oblivious key bit is as follows:

\begin{enumerate}
\item Rule out those combinations of values for the $k$ bits that are not consistent with the values for $\vec{p}_j$ received from Dave.
\item Divide the remaining possibilities into two sets --- those that correspond to an oblivious key bit of 0, and of 1.
\item Based on the measurement results and estimated error rates, calculate the probability that each combination of values for the $k$ bits is correct.  The set with the higher total probability determines the most likely value of the oblivious key bit.
\item Compute the probability of error in the oblivious key bit, $e_\mathrm{k}$.
\end{enumerate}

\noindent Note that Ursula can significantly reduce the computation required for error correction by performing this procedure only if almost all of the $k$ bits were measured conclusively.  In doing so, she only performs error correction if there is a possibility that the result will satisfy $e_\mathrm{k}\le t_\mathrm{U}$.

The error correcting codes used in this work are given by:

\begin{equation}
\mathbf{H}_{35.6} = \begin{bmatrix}
  1 & 0 & 0 & 0 & 0 & 1 & 0 & 0 & 0 & 0\\ 
  0 & 1 & 0 & 0 & 0 & 1 & 1 & 1 & 0 & 0\\
  0 & 0 & 1 & 0 & 0 & 1 & 1 & 0 & 1 & 0\\
  0 & 0 & 0 & 1 & 0 & 1 & 0 & 1 & 1 & 0\\
  0 & 0 & 0 & 0 & 1 & 1 & 0 & 0 & 0 & 1
\end{bmatrix}\label{eqn:H_35}
\end{equation}

\noindent for $\theta=35.6^\circ$ and

\begin{equation}
\mathbf{H}_{25} = \begin{bmatrix}
  1 & 0 & 0 & 0 & 0 & 1 & 0 & 0 & 0\\
  0 & 1 & 0 & 0 & 0 & 1 & 0 & 0 & 0\\
  0 & 0 & 1 & 0 & 0 & 1 & 1 & 1 & 0\\
  0 & 0 & 0 & 1 & 0 & 1 & 1 & 0 & 1\\
  0 & 0 & 0 & 0 & 1 & 1 & 0 & 1 & 1
\end{bmatrix}\label{eqn:H_25}
\end{equation}

\noindent for $\theta=25^\circ$.  They were selected using an exhaustive search of potential error-correcting codes for $k\le 10$.  The probability distribution for $e_\mathrm{k}$ is computed for each code based on the parameters in Table 2 of the main text, and the selected codes provide a low probability for $e_\mathrm{k}\le t_\mathrm{D}$ (indicating a small amount of information leakage to Ursula) as well as a suitable probability for $e_\mathrm{k}\le t_\mathrm{U}$ (ensuring that Ursula learns a few bits of the oblivious key on average).  Note that both matrices are in reduced row echelon form (i.e. no 1's appear below the leftmost 1 in any row).  This is due to the fact that the possible $k$-bit vectors remaining after step 1 of the error correction process (i.e. consistent with the parity information received from Dave) are given by the possible solutions of the system of linear equations in Eq.~\ref{eqn:parity}, hence any error correction codes that have the same reduced row echelon form behave identically in the error correction process.  The search space was thus limited by only considering matrices in reduced row echelon form.

\section{Requirements for security}

The security of the experimental results presented in Table 3 and Figure 3 of the main text hold given that the dishonest party is limited to non-quantum attacks (e.g. an arbitrarily powerful classical computer, which would be sufficient to break computational protocols using classical information such as~\cite{Rabin81}).  Furthermore, results for the security of the protocol against several quantum attacks are presented in the Section~\ref{sec:cheating}.  Note that these limitations on the attacks a dishonest party can perform are a result of the current security analysis of the protocol, and may not be required in general.  It remains an open question as to what limitations on the dishonest party, if any, are required to achieve a sufficient level of security.  Based on the attacks we have studied, we believe that at a fundamental level, the security of the protocol stems from the complementarity principle (protecting the user's security) and the superposition principle (protecting the database's security).  In addition, we note that the error-correcting code in our protocol can be selected in order to provide less information to Ursula in order to compensate for an increased information gain from more powerful quantum measurements.  Thus, it may be possible to adopt such measurements as the legitimate procedure for the user, provided that the measurements are feasible technologically.

We also note that the security results are valid only if certain requirements are met.  These requirements are listed below, beginning with those that are required in general, followed by those that are imposed by the current security analysis:\\

\begin{enumerate}
\item Ursula's and Dave's laboratories are secure (i.e. no information leaves their laboratories except as specified in the protocol).  (Required for any protocol.)
\item Quantum theory is correct and complete.  (Required for any quantum protocol.)
\item The dishonest party is limited to the attacks covered in the current security analysis (see Section~\ref{sec:cheating}).
\item In our experimental demonstration, it is also necessary to assume that the user is not able to take advantage of multi-photon pulses that result from using a source of weak coherent pulses.  While this assumption can be avoided if Dave uses a single photon source, the implementation of weak coherent pulses is much simpler from a technological perspective.  Thus, it is desirable for the protocol to be secure for weak coherent pulses without the need for additional assumptions.  The decoy state techniques used in QKD~\cite{Hwang03,Wang05,Lo05} provide security against an adversary capable of exploiting multi-photon pulses.  However, these techniques cannot be directly applied in cases where the two parties are adversarial, as is the case in private queries, and must be modified to account for the fact that the two parties need not be honest in the protocol~\cite{Wehner10}.  However, it is not clear that the techniques in ref.~\citenum{Wehner10} can be applied directly to our protocol. In particular, Ursula may gain an advantage by manipulating the aggregate statistics of the decoy state protocol by conducting an attack (e.g. by lying about detections) during a subset of the protocol while acting honestly for the remaining subset.  Analyzing and adapting decoy state techniques for our protocol is thus an interesting open question.  It may also be possible for Dave to base his estimate of the additional information that may have been extracted from multi-photon pulses on a characterization of his source.  Regardless of how Dave quantifies Ursula's information gain due to multi-photon pulses it can be accounted for by selecting an appropriate error-correcting code.  If the information gain is sufficiently small, the protocol can provide a suitable level of database security while maintaining a high success probability for the user.

\end{enumerate}

\section{Cheating Strategies\label{sec:cheating}}

In this section we discuss the attacks on individual qubits proposed in~\cite{Jakobi11, Gao11}.  The discussion below shows that the error correction step provides improved security for the protocol against these individual attacks.  Optimization of error correction in view of coherent attacks remains an interesting open question, as does an analysis of fully general quantum attacks and an information theoretic treatment of our protocol.  Furthermore, we comment on the issue of error rate estimation between adversarial parties.  As example cases for these discussions, we consider the mean parameters ($\theta$, $p_\mathrm{c}$, $e_\mathrm{c}$, and $e_\mathrm{i}$) measured with $\mu = 0.95 \pm 0.47$ using standard detectors and the simulated parameters for low-noise detectors (see Table~2 in the main text).  For the measured parameters, we do not consider the observed variances since they are specific to the system used to implement the honest protocol.\\

\subsection{User Privacy}

Let us first consider an attempt by the database to determine which piece of information Ursula is interested in.  Recall that our protocol does not prevent a dishonest database from gaining some information about Ursula's query, but is cheat sensitive in that it gives Ursula the possibility of detecting such an attack.  Performing the attack described below does not require any additional technology, as it simply requires Dave to send quantum states that either maximize or minimize the probability, $p_\mathrm{c}$, that Ursula will believe her measurement was conclusive~\cite{Jakobi11}.  In order to determine Ursula's query, Dave seeks to have Ursula learn only a single bit of the oblivious key whose position is known to him, thus he maximizes $p_\mathrm{c}$ for the $k$ bits that form one oblivious key bit in an attempt to convince Ursula that she knows a particular bit of the oblivious key.  He then minimizes $p_\mathrm{c}$ elsewhere in an attempt to prevent Ursula from knowing other bits in the oblivious key, in positions unknown to him.  As noted in~\cite{Gao11}, Dave's ability to control $p_\mathrm{c}$ improves as the angle between the 0-basis and 1-basis, $\theta$, is decreased, making the attack more powerful.  However, in both cases (i.e. maximization or minimization of $p_\mathrm{c}$), the quantum state Dave sends for this attack lies directly between either pair of $\psi$ or $\phi$ states shown in Figure~\ref{fig:SI_states}, and thus Ursula will associate a bit value to the measurement that is completely unknown to Dave.  Hence, under this attack, Ursula receives a random bit value in response to her query, leading to the cheat sensitive property in~\cite{Jakobi11, Gao11} (and in our protocol), in which incorrect query results will reveal Dave's dishonest behavior (i.e. over time, Dave will acquire a reputation of providing poor query results).

Furthermore, in our protocol the error correction steps provide additional opportunities for Ursula to verify Dave's honesty, both weakening the above attack as well as providing the possibility of detecting the weakened attack prior to Ursula revealing information about her query.  Specifically, the consequence of Dave sending quantum states that minimize $p_\mathrm{c}$ (in order to prevent Ursula from knowing one or more bits of the oblivious key in random positions) is that Ursula's and Dave's sifted keys are completely uncorrelated (i.e. they have error rates $e_\mathrm{c} = e_\mathrm{i} = 50\%$).  Additionally, since Dave has no knowledge of Ursula's sifted key, the parity bits, $\vec{p}_j$ (see Eq.~\ref{eqn:parity}), that he sends for error correction will be completely uncorrelated with the parity bits Ursula computes from her measurement results.  This allows Ursula to detect a cheating database, and abort the protocol.  While this severely restricts Dave's ability to ensure that Ursula does not know bits of the oblivious key in random positions, it does not prevent him from attempting to convince Ursula that she knows a bit in a particular position of his choosing in addition to any bits she learns randomly (in this case, Dave is unsure if Ursula's query corresponds to the position where he conducted the attack, or to an unknown position that Ursula learned randomly).  This is because Dave only needs to maximize $p_\mathrm{c}$ in $k$ bits out of $kN$ bits of the sifted key, which has a negligible effect on the overall error rates for large $N$.  However, this attack has a limited success probability, and if it fails, it may fail in a way that is suspicious to Ursula, again allowing Ursula to abort the protocol (see below for a detailed example).  Note that the above verifications occur after the error correction step, but before the shift value is communicated, thus Dave gains no information about Ursula's query if the protocol is aborted.

To illustrate the possibility for Ursula to detect an attempt by Dave to convince her that she knows a particular bit, we consider the parameters  discussed above.  For $k=10$ and $\theta=35.6^\circ$, there is a 37.49\% chance that Ursula will believe all $k$ bits are conclusive given this attack.  For $k=9$ and $\theta=25^\circ$, this probability increases to 64.93\%.  However, for Dave to convince Ursula that she knows a particular bit of the oblivious key, it is not sufficient for her to believe that all $k$ bits are conclusive, as the error correction procedure must also indicate that her measurement results are correct or correctable (i.e. the error correction procedure results in a error probability $e_\mathrm{k} \le t_\mathrm{U}$, where we recall that we have selected $t_\mathrm{U}=10^{-3}$ as the threshold below which Ursula considers a bit to be known).  The attack thus becomes more difficult with error correction, since the database must also send parity information to Ursula that is consistent with her measurements.  Since Dave's bit values are completely uncorrelated with Ursula's measured bit values, the parity information that Dave sends is essentially random, and Ursula is unlikely to find a low value for $e_\mathrm{k}$.  In the above examples, Ursula finds $e_\mathrm{k}\le 10^{-3}$ with only 5.92\% probability (for $k=10$ and $\theta=35.6^\circ$) and 12.73\% probability (for $k=9$ and $\theta=25^\circ$), showing that this attack has a limited success probability.  In addition, the case in which Ursula believes all $k$ bits were measured conclusively is of particular interest as it is very unlikely that she will find a large probability of error in the oblivious key bit after error correction, $e_\mathrm{k}$, if the protocol was performed honestly.  However, in the above attack, Dave must send parity information that is uncorrelated with Ursula's measurement results, leading to a large amount of uncertainty during Ursula's error correction process and resulting in a high probability of finding a large value for $e_\mathrm{k}$.  For example, when Ursula believes all $k$ bits were measured conclusively, for $k=10$ and $\theta=35.6^\circ$, she expects $e_\mathrm{k} \ge 0.15$ with 2.14\% probability if Dave is honest, but this value increases to 40.63\% given the above attack.  For $k=9$ and $\theta=25^\circ$, she expects $e_\mathrm{k} \ge 0.055$ with 0.71\% probability when honest, and 65.63\% with the attack.  A large value for $e_\mathrm{k}$ if all $k$ bits are measured conclusively can thus serve as an indication that Dave is attempting to cheat, and allows Ursula to abort the protocol.  Furthermore, even if the protocol proceeds and Dave is cheating (e.g. because Dave, by chance, sent consistent parity information), Ursula's and Dave's oblivious key bits after error correction are still uncorrelated, as in the protocol of~\cite{Jakobi11, Gao11}.  This ensures that the cheat sensitive property of the protocols in~\cite{Jakobi11, Gao11} discussed above is preserved in our protocol.

Generally speaking, we note that the additional benefits provided by the error correction procedure are relevant to other attack strategies as well.  Ursula now has the ability to monitor the aggregate error rates in the system, allowing her to detect any attack by Dave that has a significant effect on the overall error rates.  Furthermore, the need for the database to be able to send meaningful parity information during error correction provides an additional hurdle for attacks that cause Dave to lose information about Ursula's measurement results.\\

\subsection{Database Privacy}

On the other hand, a user attacking the protocol seeks to learn as many bits from the database as possible.  One method of doing so is to store the photons from Dave in a quantum memory until after he reveals whether he sent a $\psi$ or $\phi$ state, and then perform an unambiguous state discrimination (USD) measurement~\cite{Herzog05, Raynal06} to distinguish which of the two remaining states was sent.  However, as Dave only reveals information about a quantum state after Ursula has declared that a photon has been detected, every photon that a dishonest Ursula declares as ``detected" contributes to her sifted key.  As such, any photon that Ursula declares as ``detected'' but subsequently fails to detect (e.g. because she could not identify when a photon was successfully stored in her quantum memory, or because of loss occurring after the declaration) results in bits in the sifted key of which Ursula has no knowledge.  Successfully performing an USD attack thus requires a heralding signal indicating that a photon was successfully stored in the quantum memory, and the ability to recall the photon from the quantum memory with near 100\% efficiency.  For the following analysis, we assume a heralding signal in conjunction with a perfect quantum memory (i.e. one that introduces no error into the quantum states, and has 100\% efficiency; a realistic quantum memory, such as those assumed in the noisy-storage model, would reduce the effectiveness of the attack), and that there are no other sources of loss that reduce the success probability of the USD measurement.  

If Ursula is able to perform an USD measurement, this allows her to maximize the probability that the quantum measurements will give conclusive results.  As shown in~\cite{Jakobi11}, the probability of conclusive results increases only slightly when performing USD measurements, resulting in the user only learning a few more bits than when making honest measurements.  Furthermore, the advantage decreases as $\theta$ is decreased~\cite{Gao11}.  Additionally, in the presence of error correction, the advantage of performing an USD measurement further decreases.  This is because the USD measurement gains no information from inconclusive results, essentially exchanging this information for an increased probability of obtaining a conclusive result.  However, the partial information from inconclusive results is useful during error correction, and can even allow Ursula to know the value of the oblivious key bit in some instances in which not all measurements were conclusive.  As such, error correction can reduce the effectiveness of the USD attack.  Performing USD measurements when using the code with $k=10$ and $\theta=35.6^\circ$ only increases the average number of bits the user knows from $\bar{n}=3.89$ to $\bar{n}=11.15$ --- a rather small gain for a database of $10^6$ bits.  For the code using $k=9$ and $\theta=25^\circ$, performing USD measurements actually decreases the average number of bits the user knows from $\bar{n} = 4.35$ to $\bar{n} = 1.00$.  This decrease is due to the fact that at this smaller value of $\theta$, the value of the partial information gained from inconclusive measurements outweighs the slightly improved probability for a conclusive measurement offered by the USD measurement.  Note that these results are based on having the same error rate as for the honest measurements, which may not be a realistic assumption given that a different measurement apparatus is required.  The issue of error rates differing from those used to select the error-correcting code is addressed separately below so as to isolate this effect from that of the USD measurement.\\

\begin{table}
\centering
\caption{Comparison of simulation results for a user experiencing higher error rates than those used by Dave to select an error-correcting code.  The columns labeled  ``all'' show experimental results obtained using standard detectors ($\theta = 35.6^\circ$, $k=10$), or simulation results with improved detectors ($\theta = 25 ^\circ$, $k=9$), as taken from Tables 2 and 3 of the main text, and represent the actual results of the protocol as influenced by noise due to all imperfections.  The columns labeled ``source only'' represent Dave's predicted results for the protocol, based on an error rate estimation considering only noise introduced by his source.\label{tab:params_ec}}
\begin{tabular}{|c|c|c|c|c|}
\hline
~ & \multicolumn{2}{c|}{$\theta = 35.6^\circ$, $k=10$} & \multicolumn{2}{c|}{$\theta = 25 ^\circ$, $k=9$} \\
\hline
noise & all & source only & all & source only \\
\hline
$p_\mathrm{c}$ (\%) & 16.1 & 15.9 & 9.22 & 9.14 \\
$e_\mathrm{c}$ (\%) & 4.4 & 2.5 & 1.91 & 1.38 \\
$e_\mathrm{i}$ (\%) & 41.24 & 40.89 & 45.12 & 45.11 \\
\hline
$\bar{n}$ (bits) & 3.89 & 14.32 & 4.35 & 10.67 \\
$\bar{m}$ (\%) & 6.03 & 6.69 & 0.96 & 0.93 \\
\hline
\end{tabular}
\end{table}

\subsection{Error rate estimation}

Finally, since Ursula and Dave have an adversarial nature in the private query protocol, accurately characterizing the error rate in the system in order to select an error-correcting code is not straightforward.  In particular, Ursula would like the database to believe that the error rate is higher than in reality, as Dave would then select an error-correcting code that gives her more information, allowing her to learn more bits from the database.  To avoid this problem, Dave can determine the amount of information a user will learn from the protocol based solely on the error introduced by devices directly under his control.  In fact, he can even choose to deliberately introduce additional noise in order to provide the desired level of database security.  Additional imperfections in the system would cause the user to experience a higher error rate than Dave's estimate, leading to her learning fewer bits than the database predicts.  To show that there is a regime that allows the protocol to succeed from the user's perspective while still providing good database security, we re-examine the error-correcting codes that we have considered thus far using the parameters shown in the columns labeled ``source only'' in Table~\ref{tab:params_ec}, where noise in the system has been reduced compared to the original parameters in the main text (shown in the columns labeled ``all'').  Note that the effect of the lower noise observed by the database is not just a lower error rate in the conclusive measurements, $e_\mathrm{c}$, in the ``source only'' columns --- the other parameters are affected as well.  The error rate for inconclusive measurements, $e_\mathrm{i}$, is affected by the same noise sources as $e_\mathrm{c}$, but the effect on $e_\mathrm{i}$ is smaller as the error for inconclusive measurements is dominated by uncertainty inherent in the quantum measurement.  Hence, $e_\mathrm{i}$ in the ``source only'' columns is only slightly lower than in the ``user'' columns.  The total number of conclusive results is reduced slightly as the number of conclusive results recorded due to noise events is lower.  Hence, the probability of conclusive measurements, $p_\mathrm{c}$, is lowered slightly in the ``source only'' columns.  Table~\ref{tab:params_ec} also shows the results for the average number of bits learned by the user, $\bar{n}$, and the average proportion of the database for whic Dave considers Ursula to have significant partial information, $\bar{m}$, for the original parameters in the ``user'' columns, as well as for a lower error rate that can be used to select the error-correcting code in the ``source only'' columns.  As can be seen, the reduction in error rates does not result in a large increase in the potential amount of information gained by a user who experiences no additional error.  Thus, it is possible for an error-correcting code that is selected based on local error rates to both provide the database with good security and allow the protocol to be successful for a user experiencing higher error rates.

\bibliographystyle{naturemag}
\bibliography{ref}

\begin{thebibliography}{10}
\expandafter\ifx\csname url\endcsname\relax
  \def\url#1{\texttt{#1}}\fi
\expandafter\ifx\csname urlprefix\endcsname\relax\def\urlprefix{URL }\fi
\providecommand{\bibinfo}[2]{#2}
\providecommand{\eprint}[2][]{\url{#2}}

\bibitem{Bennett84}
\bibinfo{author}{Bennett, C.~H.} \& \bibinfo{author}{Brassard, G.}
\newblock \bibinfo{title}{Quantum cryptography: Public key distribution and
  coin tossing}.
\newblock \emph{\bibinfo{journal}{Proceedings of the IEEE International
  Conference on Computers, Systems and Signal Processing}}
  \bibinfo{pages}{175--179} (\bibinfo{year}{1984}).

\bibitem{Gisin02}
\bibinfo{author}{Gisin, N.}, \bibinfo{author}{Ribordy, G.},
  \bibinfo{author}{Tittel, W.} \& \bibinfo{author}{Zbinden, H.}
\newblock \bibinfo{title}{Quantum cryptography}.
\newblock \emph{\bibinfo{journal}{Rev. Mod. Phys.}}
  \textbf{\bibinfo{volume}{74}}, \bibinfo{pages}{145--195}
  (\bibinfo{year}{2002}).

\bibitem{Scarani09}
\bibinfo{author}{Scarani, V.} \emph{et~al.}
\newblock \bibinfo{title}{The security of practical quantum key distribution}.
\newblock \emph{\bibinfo{journal}{Rev. Mod. Phys.}}
  \textbf{\bibinfo{volume}{81}}, \bibinfo{pages}{1301--1350}
  (\bibinfo{year}{2009}).

\bibitem{Hillery99}
\bibinfo{author}{Hillery, M.}, \bibinfo{author}{Bu\ifmmode~\check{z}\else
  \v{z}\fi{}ek, V.} \& \bibinfo{author}{Berthiaume, A.}
\newblock \bibinfo{title}{Quantum secret sharing}.
\newblock \emph{\bibinfo{journal}{Phys. Rev. A}} \textbf{\bibinfo{volume}{59}},
  \bibinfo{pages}{1829--1834} (\bibinfo{year}{1999}).

\bibitem{Tittel01}
\bibinfo{author}{Tittel, W.}, \bibinfo{author}{Zbinden, H.} \&
  \bibinfo{author}{Gisin, N.}
\newblock \bibinfo{title}{Experimental demonstration of quantum secret
  sharing}.
\newblock \emph{\bibinfo{journal}{Phys. Rev. A}} \textbf{\bibinfo{volume}{63}},
  \bibinfo{pages}{042301} (\bibinfo{year}{2001}).

\bibitem{Aharonov00}
\bibinfo{author}{Aharonov, D.}, \bibinfo{author}{Ta-Shma, A.},
  \bibinfo{author}{Vazirani, U.~V.} \& \bibinfo{author}{Yao, A.~C.}
\newblock \bibinfo{title}{Quantum bit escrow}.
\newblock In \emph{\bibinfo{booktitle}{Proceedings of the thirty-second annual
  ACM symposium on Theory of computing}}, STOC '00, \bibinfo{pages}{705--714}
  (\bibinfo{year}{2000}).

\bibitem{Berlin11}
\bibinfo{author}{Berl\'{i}n, G.} \emph{et~al.}
\newblock \bibinfo{title}{Experimental loss tolerant quantum coin flipping}.
\newblock \emph{\bibinfo{journal}{Nat. Commun.}} \textbf{\bibinfo{volume}{2}},
  \bibinfo{pages}{561} (\bibinfo{year}{2011}).

\bibitem{Ng12}
\bibinfo{author}{Ng, N. H.~Y.}, \bibinfo{author}{Joshi, S.~K.},
  \bibinfo{author}{Ming, C.~C.}, \bibinfo{author}{Kurtsiefer, C.} \&
  \bibinfo{author}{Wehner, S.}
\newblock \bibinfo{title}{Experimental implementation of bit commitment in the
  noisy-storage model}.
\newblock \emph{\bibinfo{journal}{Nat. Commun.}} \textbf{\bibinfo{volume}{3}},
  \bibinfo{pages}{1326} (\bibinfo{year}{2012}).

\bibitem{Konig12}
\bibinfo{author}{K\"{o}nig, R.}, \bibinfo{author}{Wehner, S.} \&
  \bibinfo{author}{Wullschleger, J.}
\newblock \bibinfo{title}{Unconditional security from noisy quantum storage}.
\newblock \emph{\bibinfo{journal}{Information Theory, IEEE Transactions on}}
  \textbf{\bibinfo{volume}{58}}, \bibinfo{pages}{1962 --1984}
  (\bibinfo{year}{2012}).

\bibitem{Giovannetti08}
\bibinfo{author}{Giovannetti, V.}, \bibinfo{author}{Lloyd, S.} \&
  \bibinfo{author}{Maccone, L.}
\newblock \bibinfo{title}{Quantum private queries}.
\newblock \emph{\bibinfo{journal}{Phys. Rev. Lett.}}
  \textbf{\bibinfo{volume}{100}}, \bibinfo{pages}{230502}
  (\bibinfo{year}{2008}).

\bibitem{DeMartini09}
\bibinfo{author}{De~Martini, F.} \emph{et~al.}
\newblock \bibinfo{title}{Experimental quantum private queries with linear
  optics}.
\newblock \emph{\bibinfo{journal}{Phys. Rev. A}} \textbf{\bibinfo{volume}{80}},
  \bibinfo{pages}{010302} (\bibinfo{year}{2009}).

\bibitem{Jakobi11}
\bibinfo{author}{Jakobi, M.} \emph{et~al.}
\newblock \bibinfo{title}{Practical private database queries based on a
  quantum-key-distribution protocol}.
\newblock \emph{\bibinfo{journal}{Phys. Rev. A}} \textbf{\bibinfo{volume}{83}},
  \bibinfo{pages}{022301} (\bibinfo{year}{2011}).

\bibitem{Gao11}
\bibinfo{author}{Gao, F.}, \bibinfo{author}{Liu, B.}, \bibinfo{author}{Wen,
  Q.-Y.} \& \bibinfo{author}{Chen, H.}
\newblock \bibinfo{title}{Flexible quantum private queries based on quantum key
  distribution}.
\newblock \emph{\bibinfo{journal}{Opt. Express}} \textbf{\bibinfo{volume}{20}},
  \bibinfo{pages}{17411--17420} (\bibinfo{year}{2012}).

\bibitem{Lo97}
\bibinfo{author}{Lo, H.-K.}
\newblock \bibinfo{title}{Insecurity of quantum secure computations}.
\newblock \emph{\bibinfo{journal}{Phys. Rev. A}} \textbf{\bibinfo{volume}{56}},
  \bibinfo{pages}{1154--1162} (\bibinfo{year}{1997}).

\bibitem{Rabin81}
\bibinfo{author}{Rabin, M.~O.}
\newblock \bibinfo{title}{How to exchange secrets by oblivious transfer}.
\newblock \bibinfo{type}{Tech. Rep.}, \bibinfo{institution}{Harvard University}
  (\bibinfo{year}{1981}).

\bibitem{Naor00}
\bibinfo{author}{Naor, M.} \& \bibinfo{author}{Pinkas, B.}
\newblock \bibinfo{title}{Distributed oblivious transfer}.
\newblock In \emph{\bibinfo{booktitle}{Proceedings of the 6th International
  Conference on the Theory and Application of Cryptology and Information
  Security: Advances in Cryptology}}, ASIACRYPT '00, \bibinfo{pages}{205--219}
  (\bibinfo{year}{2000}).

\bibitem{Blundo07}
\bibinfo{author}{Blundo, C.}, \bibinfo{author}{D'Arco, P.},
  \bibinfo{author}{De~Santis, A.} \& \bibinfo{author}{Stinson, D.}
\newblock \bibinfo{title}{On unconditionally secure distributed oblivious
  transfer}.
\newblock \emph{\bibinfo{journal}{Journal of Cryptology}}
  \textbf{\bibinfo{volume}{20}}, \bibinfo{pages}{323--373}
  (\bibinfo{year}{2007}).

\bibitem{Erven13}
\bibinfo{author}{Erven, C.} \emph{et~al.}
\newblock \bibinfo{title}{An experimental implementation of oblivious transfer
  in the noisy storage model}.
\newblock \emph{\bibinfo{journal}{arXiv:1308.5098}}  (\bibinfo{year}{2013}).

\bibitem{Kleinjung10}
\bibinfo{author}{Kleinjung, T.} \emph{et~al.}
\newblock \bibinfo{title}{Factorization of a 768-bit {RSA} modulus}.
\newblock In \emph{\bibinfo{booktitle}{Proceedings of the 30th annual
  conference on Advances in cryptology}}, CRYPTO'10, \bibinfo{pages}{333--350}
  (\bibinfo{year}{2010}).

\bibitem{Shor97}
\bibinfo{author}{Shor, P.~W.}
\newblock \bibinfo{title}{Polynomial-time algorithms for prime factorization
  and discrete logarithms on a quantum computer}.
\newblock \emph{\bibinfo{journal}{SIAM Journal on Computing}}
  \textbf{\bibinfo{volume}{26}}, \bibinfo{pages}{1484--1509}
  (\bibinfo{year}{1997}).

\bibitem{Lvovsky09}
\bibinfo{author}{Lvovsky, A.~I.}, \bibinfo{author}{Sanders, B.~C.} \&
  \bibinfo{author}{Tittel, W.}
\newblock \bibinfo{title}{Optical quantum memory}.
\newblock \emph{\bibinfo{journal}{Nat. Photon.}} \textbf{\bibinfo{volume}{3}},
  \bibinfo{pages}{706--714} (\bibinfo{year}{2009}).

\bibitem{Tittel10}
\bibinfo{author}{Tittel, W.} \emph{et~al.}
\newblock \bibinfo{title}{Photon-echo quantum memory in solid state systems}.
\newblock \emph{\bibinfo{journal}{Laser \& Photonics Reviews}}
  \textbf{\bibinfo{volume}{4}}, \bibinfo{pages}{244--267}
  (\bibinfo{year}{2010}).

\bibitem{Hammerer10}
\bibinfo{author}{Hammerer, K.}, \bibinfo{author}{S\o{}rensen, A.~S.} \&
  \bibinfo{author}{Polzik, E.~S.}
\newblock \bibinfo{title}{Quantum interface between light and atomic
  ensembles}.
\newblock \emph{\bibinfo{journal}{Rev. Mod. Phys.}}
  \textbf{\bibinfo{volume}{82}}, \bibinfo{pages}{1041--1093}
  (\bibinfo{year}{2010}).

\bibitem{Simon10}
\bibinfo{author}{Simon, C.} \emph{et~al.}
\newblock \bibinfo{title}{Quantum memories}.
\newblock \emph{\bibinfo{journal}{The European Physical Journal D}}
  \textbf{\bibinfo{volume}{58}}, \bibinfo{pages}{1--22} (\bibinfo{year}{2010}).

\bibitem{Schindler11}
\bibinfo{author}{Schindler, P.} \emph{et~al.}
\newblock \bibinfo{title}{Experimental repetitive quantum error correction}.
\newblock \emph{\bibinfo{journal}{Science}} \textbf{\bibinfo{volume}{332}},
  \bibinfo{pages}{1059--1061} (\bibinfo{year}{2011}).

\bibitem{Bussieres13}
\bibinfo{author}{Bussi\`{e}res, F.} \emph{et~al.}
\newblock \bibinfo{title}{Prospective applications of optical quantum
  memories}.
\newblock \emph{\bibinfo{journal}{arXiv:1306.6904}}  (\bibinfo{year}{2013}).

\bibitem{Scarani04}
\bibinfo{author}{Scarani, V.}, \bibinfo{author}{Ac\'{\i}n, A.},
  \bibinfo{author}{Ribordy, G.} \& \bibinfo{author}{Gisin, N.}
\newblock \bibinfo{title}{Quantum cryptography protocols robust against photon
  number splitting attacks for weak laser pulse implementations}.
\newblock \emph{\bibinfo{journal}{Phys. Rev. Lett.}}
  \textbf{\bibinfo{volume}{92}}, \bibinfo{pages}{057901}
  (\bibinfo{year}{2004}).

\bibitem{Vernam26}
\bibinfo{author}{Vernam, G.~S.}
\newblock \bibinfo{title}{Cipher printing telegraph systems for secret wire and
  radio telegraphic communications}.
\newblock \emph{\bibinfo{journal}{American Institute of Electrical Engineers,
  Transactions of the}} \textbf{\bibinfo{volume}{XLV}}, \bibinfo{pages}{295
  --301} (\bibinfo{year}{1926}).

\bibitem{MacKay03}
\bibinfo{author}{MacKay, D.}
\newblock \emph{\bibinfo{title}{Information Theory, Inference, and Learning
  Algorithms}} (\bibinfo{publisher}{Cambridge University Press},
  \bibinfo{year}{2003}).

\bibitem{Herzog05}
\bibinfo{author}{Herzog, U.} \& \bibinfo{author}{Bergou, J.~A.}
\newblock \bibinfo{title}{Optimum unambiguous discrimination of two mixed
  quantum states}.
\newblock \emph{\bibinfo{journal}{Phys. Rev. A}} \textbf{\bibinfo{volume}{71}},
  \bibinfo{pages}{050301} (\bibinfo{year}{2005}).

\bibitem{Raynal06}
\bibinfo{author}{Raynal, P.}
\newblock \bibinfo{title}{Unambiguous state discrimination of two density
  matrices in quantum information theory}.
\newblock \emph{\bibinfo{journal}{arXiv:quant-ph/0611133v1}}
  (\bibinfo{year}{2006}).

\bibitem{Lucio09}
\bibinfo{author}{Lucio-Martinez, I.}, \bibinfo{author}{Chan, P.},
  \bibinfo{author}{Mo, X.-F.}, \bibinfo{author}{Hosier, S.} \&
  \bibinfo{author}{Tittel, W.}
\newblock \bibinfo{title}{{P}roof-of-concept of real world quantum key
  distribution with quantum frames}.
\newblock \emph{\bibinfo{journal}{New J. Phys.}} \textbf{\bibinfo{volume}{11}},
  \bibinfo{pages}{095001} (\bibinfo{year}{2009}).

\bibitem{Hwang03}
\bibinfo{author}{Hwang, W.-Y.}
\newblock \bibinfo{title}{Quantum key distribution with high loss: Toward
  global secure communication}.
\newblock \emph{\bibinfo{journal}{Phys. Rev. Lett.}}
  \textbf{\bibinfo{volume}{91}}, \bibinfo{pages}{057901}
  (\bibinfo{year}{2003}).

\bibitem{Wang05}
\bibinfo{author}{Wang, X.-B.}
\newblock \bibinfo{title}{Beating the photon-number-splitting attack in
  practical quantum cryptography}.
\newblock \emph{\bibinfo{journal}{Phys. Rev. Lett.}}
  \textbf{\bibinfo{volume}{94}}, \bibinfo{pages}{230503}
  (\bibinfo{year}{2005}).

\bibitem{Lo05}
\bibinfo{author}{Ma, X.}, \bibinfo{author}{Qi, B.}, \bibinfo{author}{Zhao, Y.}
  \& \bibinfo{author}{Lo, H.-K.}
\newblock \bibinfo{title}{Practical decoy state for quantum key distribution}.
\newblock \emph{\bibinfo{journal}{Phys. Rev. A}} \textbf{\bibinfo{volume}{72}},
  \bibinfo{pages}{012326} (\bibinfo{year}{2005}).

\bibitem{Wehner10}
\bibinfo{author}{Wehner, S.}, \bibinfo{author}{Curty, M.},
  \bibinfo{author}{Schaffner, C.} \& \bibinfo{author}{Lo, H.-K.}
\newblock \bibinfo{title}{Implementation of two-party protocols in the
  noisy-storage model}.
\newblock \emph{\bibinfo{journal}{Phys. Rev. A}} \textbf{\bibinfo{volume}{81}},
  \bibinfo{pages}{052336} (\bibinfo{year}{2010}).

\bibitem{Marsili13}
\bibinfo{author}{Marsili, F.} \emph{et~al.}
\newblock \bibinfo{title}{Detecting single infrared photons with 93\% system
  efficiency}.
\newblock \emph{\bibinfo{journal}{Nat. Photon.}} \textbf{\bibinfo{volume}{7}},
  \bibinfo{pages}{210--214} (\bibinfo{year}{2013}).

\bibitem{Yan12}
\bibinfo{author}{Yan, Z.} \emph{et~al.}
\newblock \bibinfo{title}{An ultra low noise telecom wavelength free running
  single photon detector using negative feedback avalanche diode}.
\newblock \emph{\bibinfo{journal}{Review of Scientific Instruments}}
  \textbf{\bibinfo{volume}{83}}, \bibinfo{pages}{073105 --073105--15}
  (\bibinfo{year}{2012}).

\bibitem{Dixon08}
\bibinfo{author}{Dixon, A.~R.}, \bibinfo{author}{Yuan, Z.~L.},
  \bibinfo{author}{Dynes, J.~F.}, \bibinfo{author}{Sharpe, A.~W.} \&
  \bibinfo{author}{Shields, A.~J.}
\newblock \bibinfo{title}{Gigahertz decoy quantum key distribution with 1
  {M}bit/s secure key rate}.
\newblock \emph{\bibinfo{journal}{Opt. Express}} \textbf{\bibinfo{volume}{16}},
  \bibinfo{pages}{18790--18979} (\bibinfo{year}{2008}).

\bibitem{PandurangaRao13}
\bibinfo{author}{Panduranga~Rao, M.~V.} \& \bibinfo{author}{Jakobi, M.}
\newblock \bibinfo{title}{Towards communication-efficient quantum oblivious key
  distribution}.
\newblock \emph{\bibinfo{journal}{Phys. Rev. A}} \textbf{\bibinfo{volume}{87}},
  \bibinfo{pages}{012331} (\bibinfo{year}{2013}).

\end{thebibliography}

\end{document}